\documentclass{PoS}

\title{Opening angles and shapes of parsec-scale \\AGN jets}

\ShortTitle{Opening angles and shapes of parsec-scale AGN jets}

\author{\speaker{Alexander~B.~Pushkarev}%
         \thanks{{\bf Acknowledgments.} This research has made use of data from the MOJAVE database 
         that is maintained by the MOJAVE team. The MOJAVE project was supported under NASA-Fermi 
         grants NNX12A087G and 11-Fermi11-0019. The National Radio Astronomy Observatory is a facility 
         of the National Science Foundation operated under cooperative agreement by Associated Universities, 
         Inc. This study was supported in part by the Russian Foundation for Basic Research grant 13-02-12103.
}\\
        Pulkovo Observatory, Pulkovskoe Chaussee 65/1, St. Petersburg 196140, Russia\\
        Crimean Astrophysical Observatory, Nauchny 298409, Crimea, Russia\\
        Max-Planck-Institut f\"ur Radioastronomie, Auf dem H\"ugel 69, 53123 Bonn, Germany \\
        E-mail: \email{pushkarev.alexander@gmail.com}}

\author{Matthew~L.~Lister\\
        Department of Physics, 
        Purdue University, 525 Northwestern Avenue, West Lafayette, IN 47907, USA}

\author{Yuri~Y.~Kovalev\\
        Astro Space Center of Lebedev Physical Institute, Profsoyuznaya 84/32, Moscow 117997, Russia\\
        Max-Planck-Institut f\"ur Radioastronomie, Auf dem H\"ugel 69, 53123 Bonn, Germany
        }

\author{Tuomas Savolainen\\
        Max-Planck-Institut f\"ur Radioastronomie, Auf dem H\"ugel 69, 53123 Bonn, Germany\\
        Aalto University Mets\"ahovi Radio Observatory, Mets\"ahovintie 114, 02540 Kylm\"al\"a, Finland
}

\abstract{We used 15 GHz VLBA observations of 366 sources having at least 5 epochs within
a time interval 1995--2013 from the \mbox{MOJAVE} program and/or its predecessor, the 2~cm 
VLBA Survey. For each source we produced a corresponding stacked image averaging all 
available epochs for a better reconstruction of the cross section of the flow. We have 
analyzed jet profiles transverse to the local jet ridge line and derived both apparent 
and intrinsic opening angles of the parsec-scale outflows.
The sources detected by the {\it Fermi}\, Large Area Telescope (LAT) during the first 
24 months of operation show wider apparent jet opening angle and smaller viewing angles 
on a very high level of significance supporting our early findings.
Analyzing transverse shapes of the outflows we found that most sources have 
conical jet geometry at parsec scales, though there are also sources that exhibit active
jet collimation.
}

\FullConference{12th European VLBI Network Symposium and Users Meeting - EVN 2014,\\
                7-10 October, 2014\\
                Cagliari, Italy}

\begin{document}

\section{Apparent opening angles}
\begin{figure}
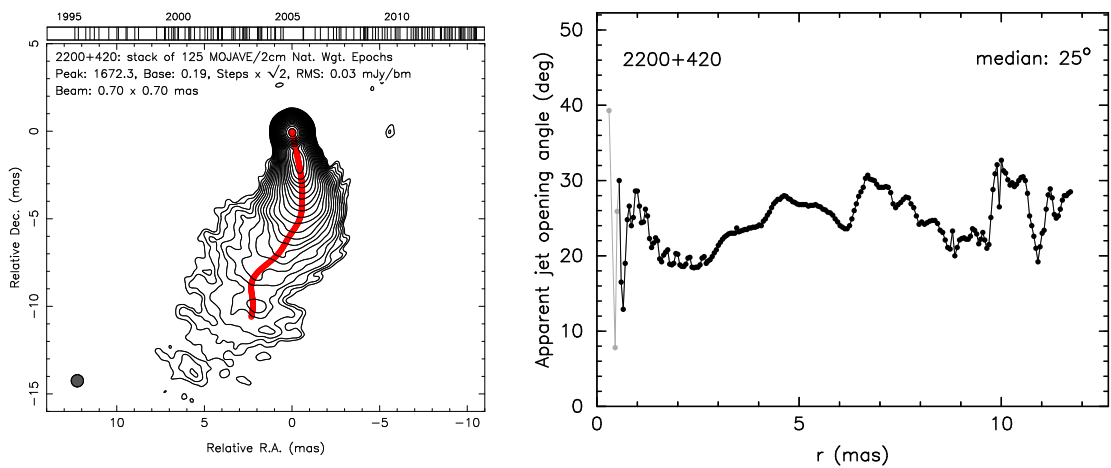

\centering
\includegraphics[height=.42\textwidth,angle=-90]{fig1a.eps}\hspace{0.5cm}
\includegraphics[height=.51\textwidth,angle=-90]{fig1b.eps}
\caption{{\it Left:} MOJAVE image of BL Lac at 15~GHz after stacking 125 epochs covering a time range of 
  about 20 years. The total intensity ridge line is shown by red. Time evolution of the ridge line in BL Lac
  is discussed in \cite{Cohen_BLLAC2}. {\it Right:} Apparent opening angle of the jet along its ridge line.}
\label{fig1}
\end{figure}

Recently we found that nearly all of the 60 heavily observed jets display significant changes of the innermost
jet position angle with time \cite{MOJAVEX} suggesting that the superluminal AGN jet features seen in a
single-epoch images occupy only a portion of the entire jet cross section. Thus, 
to better reconstruct the jet cross section of 
each source observed
within the MOJAVE/2cm VLBA Survey we produced a corresponding stacked image using all available epochs for
a given source at 15~GHz, comprising a sample of 363 AGN jets having at least 5 epochs and a clear VLBI core
position. The opening angle of the jet was calculated as the median value of
$\alpha_\mathrm{app}=2\arctan[0.5(D^2-b_\varphi^2)^{1/2}/r]$, where $D$ is the FWHM of a Gaussian fitted to
the transverse jet brightness profile, $r$ is the distance to the VLBI core along the jet ridge line,
$b_\varphi$ is the beam size along the position angle $\varphi$ of the jet-cut. 
In Fig.~\ref{fig1} we show the 15~GHz stacked image of BL Lac as an example together with opening angle of 
the jet as a function of angular distance to the core along the ridge line.

\begin{figure}
\centering
\includegraphics[height=.63\textwidth,angle=-90,clip=true]{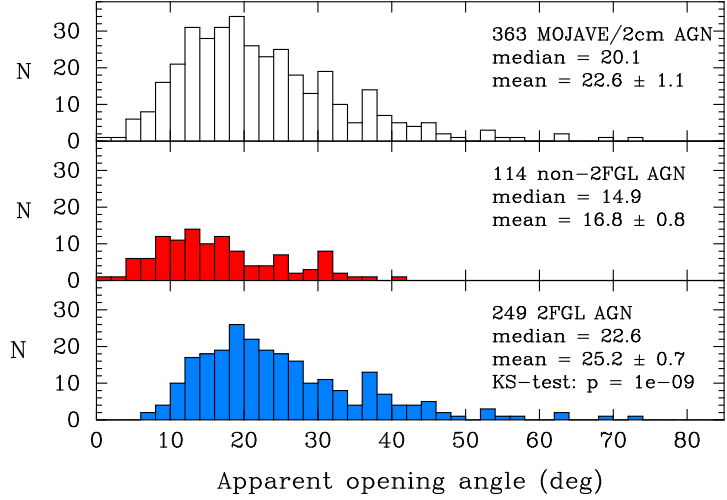}
\caption{Distributions of the apparent opening angle from jet-cut  analysis for 363 MOJAVE AGN
 ({\it top panel}), comprising 114 non-LAT-detected ({\it middle panel}) and 249 LAT-detected
 ({\it bottom panel}) sources after 24 months of scientific operation.}
\label{fig2}
\end{figure}

The distribution of the derived $\alpha_\mathrm{app}$ values is shown in Fig.~\ref{fig2} (top). LAT-detected
\cite{FGL2}
sources (Fig.~\ref{fig2}, bottom) have statistically wider apparent jet opening angles compared to those of
non-LAT-detected (Fig.~\ref{fig2}, middle) on a very high level of significance supporting our early findings.

\section{Intrinsic opening angles and viewing angles}
\begin{figure}
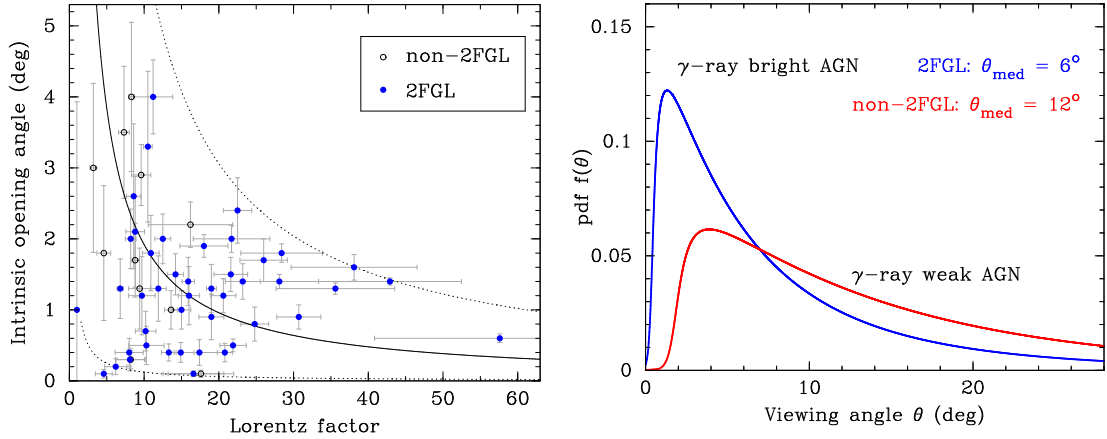

\centering
\includegraphics[height=.47\textwidth,angle=-90]{fig3a.eps}\hspace{0.3cm}
\includegraphics[height=.47\textwidth,angle=-90]{fig3b.eps}
\caption{{\it Left:} Intrinsic opening angle vs. Lorentz factor for 56 jets. The solid line shows the median
 curve fit with the assumed relation $\alpha_\mathrm{int}=\rho/\Gamma$, where $\rho$ is a constant (here
 $\rho=0.33$ rad). {\it Right:} Probability density function of viewing angle as derived from the apparent
 angle and Lorentz factor distributions of $\gamma$-ray bright (blue curve) and $\gamma$-ray weak (red curve) AGN.}
\label{fig3}
\end{figure}

We have derived the values of the viewing angle $\theta$ and the bulk Lorentz factor $\Gamma$ using 
jet speeds from the MOJAVE kinematic analysis \cite{MOJAVEX} and variability Doppler factor 
from the Mets\"ahovi AGN monitoring program \cite{Hovatta09}. The overlap of the MOJAVE and 
Mets\"ahovi programs comprises 56 sources. The intrinsic opening angles calculated for the 56 sources 
using a relation $\tan(\alpha_\mathrm{int}/2)=\tan(\alpha_\mathrm{app}/2)\sin\theta$ have a median of 
$1\hbox{$.\!\!^\circ$}3$ and show inverse dependence on Lorentz factor (Fig.~\ref{fig3}, left), as predicted by 
hydrodynamical \cite{BK79} and magnetic acceleration models \cite{Komissarov07} of relativistic jets.

A K-S test indicates no significant difference ($p=0.43$) between the samples of LAT-detected and 
non-LAT-detected sources, suggesting that the established statistical difference in apparent opening 
angles is the result of projection effects, i.e., the $\gamma$-ray bright jets are viewed at preferentially
smaller angles. Indeed, using Monte-Carlo simulation together with the Generalized Lambda Distribution
technique we have derived probability density functions of viewing angle for the LAT-detected and
non-LAT-detected sources (Fig.~\ref{fig3}, right), showing that jets of the $\gamma$-ray bright AGN tend to
have smaller angles to the line of sight comparing to those of $\gamma$-ray weak AGN, with median
values $6^\circ$ and $12^\circ$, respectively.

\section{Jet shapes}
\begin{figure}
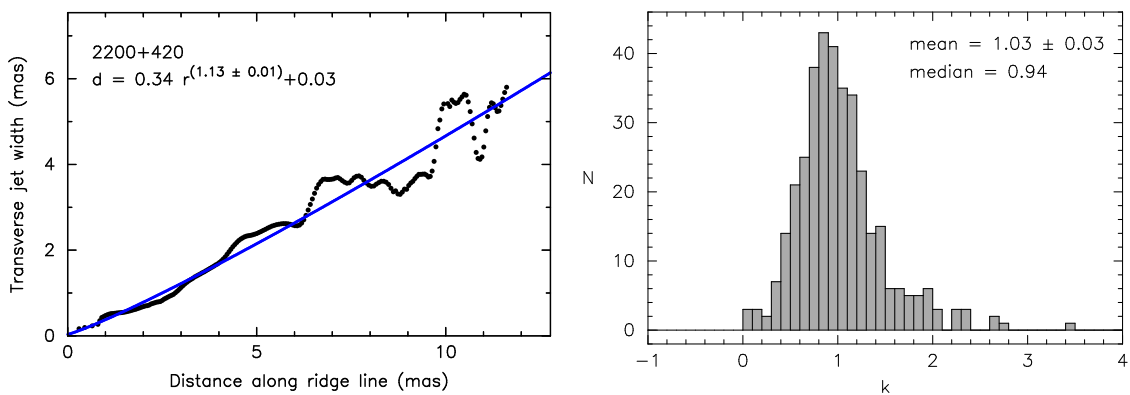

\centering
\includegraphics[height=.48\textwidth,angle=-90]{fig4a.eps}\hspace{0.3cm}
\includegraphics[height=.48\textwidth,angle=-90]{fig4b.eps}
\caption{{\it Left:} Jet width $d$ vs distance $r$ for BL Lac. The blue line is the best fit of an 
 assumed power-law dependence $d\propto r^k$. {\it Right:} Distribution of the power-law index $k$, 
 with a peak at $k\sim1$ implying that a jet shape is close to conical for most parsec-scale AGN outflows.}
\label{fig4}
\end{figure}

To study shapes of the outflows, we analyzed dependence between jet widths $d=(D^2-b^2_\varphi)^{1/2}$
derived from the profiles transverse to the local jet direction and angular separation $r$ measured
along the reconstructed total intensity ridge line. We assumed a power-law dependence $d\propto r^k$
and searched for the best fit space parameters using $\chi^2$ minimization (Fig.~\ref{fig4}, left).

The distribution of the derived power-law index $k$ presented in Fig.~\ref{fig4}, (right) peaks at values 
close to 1, suggesting that parsec-scale AGN jets typically manifest a shape close to conical. At the same time, 
a number of sources show $k$ values significantly smaller than 1, indicating that the jets undergo collimation
on scales probed by our observations (e.g., $k=0.47\pm0.01$ for M\,87). 
BL Lacs and quasars have similar $k$-distributions, with the means of $1.13\pm0.06$ and $1.01\pm0.04$,
respectively, whereas galaxies show on average smaller $k=0.83\pm0.12$, most probably because
these objects are systematically closer, and their jets are oriented at larger angles to the line of sight,
allowing us to probe the jet regions at shorter linear separations from the central engine where the outflows
become organized more effectively \cite{Hada13}.

\end{document}